\documentclass{pnastwo_mod}
\usepackage{PNAStwoF}

\usepackage{mathptmx} 
\usepackage{amsmath,mathtools}
\usepackage{calc}
\usepackage{graphicx}

\begin{document}

\title{Toward computational cumulative biology by combining models of
  biological datasets}

\author{Ali Faisal\affil{1}{Helsinki Institute for Information
    Technology HIIT, Department of Information and Computer Science,
    PO Box 15400, Aalto University, 00076 Aalto, Finland}, Jaakko
  Peltonen\affil{1}{}, Elisabeth Georgii\affil{1}{}, Johan
  Rung\affil{2}{European Molecular Biology Laboratory, European
    Bioinformatics Institute (EMBL-EBI), Wellcome Trust Genome Campus
    Hinxton, CB10 1SD, UK}\thanks{Current address: Department of Immunology, Genetics and Pathology, Science for Life Laboratory, Uppsala University, 751 85 Uppsala, Sweden.} \and Samuel Kaski\affil{1}{}\affil{3}{Helsinki Institute for Information Technology HIIT, Department of Computer Science, PO Box 68, 00014 University of Helsinki, Finland}}

\maketitle

\begin{article}

\begin{abstract} 
    A main challenge of data-driven sciences is how to make maximal
    use of the progressively expanding databases of experimental
    datasets in order to keep research cumulative. We introduce the
    idea of a modeling-based dataset retrieval engine designed for
    relating a researcher's experimental dataset to earlier work in
    the field.  The search is (i) data-driven to enable new findings,
    going beyond the state of the art of keyword searches in
    annotations, (ii) modeling-driven, to both include biological
    knowledge and insights learned from data, and (iii) scalable, as
    it is accomplished without building one unified grand model of all
    data. Assuming each dataset has been modeled beforehand, by the
    researchers or by database managers, we apply a rapidly computable
    and optimizable combination model to decompose a new dataset into
    contributions from earlier relevant models.  By using the
    data-driven decomposition we identify a network of interrelated
    datasets from a large annotated human gene expression atlas.
    While tissue type and disease were major driving forces for
    determining relevant datasets, the found relationships were richer
    and the model-based search was more accurate than keyword search;
    it moreover recovered biologically meaningful relationships that
    are not straightforwardly visible from annotations, for instance,
    between cells in different developmental stages such as thymocytes
    and T-cells.  Data-driven links and citations matched to a large
    extent; the data-driven links even uncovered corrections to the
    publication data, as two of the most linked datasets were not
    highly cited and turned out to have wrong publication entries in
    the database.
\end{abstract}

\section{Significance}

Measurement data sets of molecular biology and other experimental
sciences are being collected comprehensively to openly accessible
databases. We demonstrate that it is now possible to relate new
results to earlier science by searching with the actual data, instead
of only in the textual annotations which are restricted to known
findings and are the current state of the art. In a gene expression
database, the data-driven relationships between datasets matched well
with citations between the corresponding research papers, and even
found mistakes in the database.
\vspace{2mm}

Molecular biology, historically driven by the pursuit of
experimentally characterizing each component of the living cell, has
been transformed into a data-driven science \cite{Greene11,Tanay05,
  Caldas12, Adler09, Schmid12, Gerber07} with just as much importance
given to the computational and statistical analysis as to experimental
design and assay technology. This has brought to the fore new
computational challenges, such as the processing of massive new
sequencing data, and new statistical challenges arising from the
problem of having relatively few ($n$) samples characterized for
relatively many ($p$) variables---the ``large $p$, small $n$''
problem. High throughput technologies often are developed to assay
many parallel variables for a single sample in a run, rather than many
parallel samples for a single variable, whereas the statistical power
to infer properties of biological conditions increases with larger
sample sizes. For cost reasons, most labs are restricted to generating
datasets with the statistical power to detect only the strongest
effects. In combination with the penalties of multiple hypothesis
testing, the limitations of ``large $p$, small $n$'' datasets are
obvious. It is therefore not surprising that much work has been
devoted to address this problem.

Some of the most successful methods rely on increasing the effective
number of samples by combining with data from other, similarly
designed, experiments, in a large meta-analysis \cite{Tseng12}.
Unfortunately, this is not straightforward either. Although public
data repositories, such as the ones at NCBI in the United States and
the EBI in Europe, serve the research community with ever-growing
amounts of experimental data, they largely rely on annotation and
meta-data provided by the submitter. Database curators and semantic
tools such as ontologies provide some help in harmonizing and
standardizing the annotation, but the user who wants to find datasets
that are combinable with her own, most often must resort to searches
in free text or in controlled vocabularies which would need much
downstream curation and data analysis before any meta-analysis can be
done \cite{Rung12}.

Ideally, we would like to let the data speak for themselves. Instead
of searching for datasets that have been described similarly, which
may not correspond to a statistical similarity in the datasets
themselves, we would like to conduct that search in a data-driven way,
using the dataset itself as the query, or a statistical (rather than a
semantic) description of it. This is implicitly done for example in
multi-task learning, a method from the machine learning field
\cite{Baxter97,Caruana97}, where several related estimation tasks are
pursued together, assuming shared properties across tasks. Multi-task
learning is a form of global analysis, which builds a single unified
model of the datasets. But as the number of datasets keeps increasing
and the amount of quantitative biological knowledge keeps
accumulating, the complexity of building an accurate unified model
becomes increasingly prohibitive.

Addressing the ``large $p$, small $n$'' problem requires taking into
account both the uncertainty in the data and the existing biological
knowledge.  We now consider the hypothesized scenario where future
researchers increasingly develop hypotheses in terms of
(probabilistic) models of their data. Although far from realistic
today, a similar trend exists for sequence motif data, which are often
published as Hidden Markov models, for instance in the Pfam database
\cite{Punta12}.

In this paper we report on a feasibility study towards the scenario
where a large number of experiments have been modeled beforehand,
potentially by the researcher generating the data or the database
storing the model together with the data. We ask \emph{what could be
  done with these models towards cumulatively building knowledge from
  data in molecular biology}. Speaking about models generally and
assuming the many practical issues can be solved technically, we
arrive at our answer: \emph{a modeling-driven dataset retrieval
  engine}, which a researcher can use for positioning her own
measurement data into the context of the earlier biology. The engine
will point out relationships between experiments in the form of the
retrieval results, which is a naturally understandable interface. The
retrieval will be based on data instead of the state of the art of
using keywords and ontologies, which will make unexpected and
previously unknown findings possible. The retrieval will use the
models of the datasets which, by our assumption above, incorporate
what the researchers producing the data thought was important, but the
retrieval will be designed to be more scalable than building one
unified grand model of all data. This also implies that the way the
models are utilized needs to be approximate. Compared to existing
data-driven retrieval methods \cite{Schmid12, Caldas12}, whole
datasets, incorporating the experimental designs, will be matched
instead of individual observations. The remaining question is how to
design the retrieval so that it both reveals the interesting and
important relationships and is fast to compute.

The model we present is a first step towards this goal. We assume a
new dataset can be explained by a combination of the models for the
earlier datasets and a novelty term.  This is a mixture modeling or
regression task, in which the weights can be computed rapidly; the
resulting method scales well to large numbers of datasets, and the
speed of the mixture modeling does not depend on the sizes of the
earlier datasets.  The largest weights in the mixture model point at
the most relevant earlier datasets. The method is applicable to
several types of measurement datasets, assuming suitable models
exist. Unlike traditional mixture modeling we do not limit the form of
the mixture components, thus we bring in the knowledge built into the
stored models of each dataset.  We apply this approach to a large set
of experiments from EBI's ArrayExpress gene expression database
\cite{Lukk12}, treating each experiment in turn as a new dataset,
queried against all earlier datasets. Under our assumptions the
retrieval results can be interpreted as studies that the authors of
the study generating the query set could have cited, and we show that
the actual citations overlap with the retrieval results. The
discovered links between datasets additionally enable forming a ``hall
of fame'' of gene expression studies, containing the studies that
would have been influential assuming the retrieval system would have
existed. The links in the ``hall of fame'' verify and complement the
citation links: in our study they revealed corrections to the citation
data, as two frequently retrieved studies were not highly cited and
turned out to have erroneous publication entries in the database. We
provide an online resource for exploring and searching this ``hall of
fame'': {\tt
  http://research.ics.aalto.fi/mi/setretrieval}.

Earlier work on relating datasets has provided partial solutions along
this line, with the major limitation of being restricted to pairwise
dataset comparisons, in contrast to the proposed approach of
decomposing a dataset into contributions from a set of earlier
datasets.  Russ and Futschik \cite{Russ10} represented each dataset by
pairwise correlations of genes, and used them to compute dataset
similarities.  This dataset representation is ill-suited for typical
functional genomics experiments as a large number of samples is
required to sensibly estimate gene correlation matrices.  In addition,
it makes the dataset comparison computationally expensive, as the
representation is bulkier than the original dataset. In other works
specific case-control designs \cite{plosDisease} or known biological
processes \cite{Huttenhower08} are assumed; we generalize to
decompositions over arbitrary models.

\section{Combination of stored models for dataset retrieval}
Our goal is to infer data-driven relationships between a new ``query''
dataset $q$ and earlier datasets. The query is a dataset of $N_q$
samples $\{x_i^q\}_{i=1}^{N_q}$; in the ArrayExpress study the samples
are gene expression profiles, with the element $x_{ij}^q$ being
expression of the gene set $j$ in the sample $i$ of the query $q$, but
the setup is general and applicable to other experimental data as
well. Assume further a dataset repository of $N_S$ earlier datasets,
and assume that each dataset $s_j$, $j=1,\ldots,N_S$, has already been
modeled with a model denoted by $M^{s_j}$, later called a base model.
The base models are assumed to be probabilistic generative models,
{\it i.e.}, principled data descriptions capturing prior knowledge and
data-driven discoveries under specific distributional assumptions.
Base models for different datasets may come from different model
families, as chosen by the researchers who authored each dataset. In
this paper we use two types of base models, which are discrete
variants of principal component analysis ({\it Results}), but any
probabilistic generative models can be applied.

Assume tentatively that the dataset repository contains a library of
``base experiments'', carefully selected to induce all important known
biological effects with suitable design factors. In the special
example case of metagenomics with known constituent organisms, an
obvious set of base experiments would be the set of genomes of those
organisms \cite{Meinicke11}. A new experiment could then be expressed
as a combination of the base experiments, and potential novel
effects. More generally, for instance in a broad gene expression
atlas, it would be hard if not impossible to settle on a clean,
well-defined and up-to-date base set of experiments to correspond to
each known effect, so we choose to \emph{use the comprehensive
  collection of experiments in the current databases as the base
  experiments}. The problem setting then changes, from searching for a
unique explanation of the new experiment, to the down-to-earth and
realistic task of data-driven retrieval of a set of relevant earlier
experiments, relevant in the sense of having induced one or more of
the known or as-of-yet unknown biological effects.

We combine the earlier datasets by a method that is probabilistic but
simple and fast.  We build a \emph{combination model} for the query
dataset as a mixture model of base distributions $p(x|M^{s_j})$, which
have been estimated beforehand. In our scenario, generative models
$M^{s_j}$ are available in the repository along with datasets $s_j$;
note that the $M^{s_j}$ need not all have the same form.  In the
mixture model parameterized by $\boldsymbol{\Theta}^q =
\{\theta_j^q\}_{j=1}^{N_S + 1}$, the likelihood of observing the query
is
\begin{equation}
p(\{x_i^q\}_{i=1}^{N_q}; \boldsymbol{\Theta}^q) 
= \prod_{i=1}^{N_q} \Big[\Big(\sum_{j=1}^{N_S} \theta^q_j p(x_i^q | M^{s_j}) \Big)+ \theta^q_{N_S + 1} p(x_i^q|\psi) \Big]
\end{equation}
where $\theta^q_j$ is the mixture proportion or \emph{weight} of the
$j$th base distribution (model of dataset $s_j$) and $\theta^q_{N_S +
  1}$ is the weight for the novelty term. The novelty is modeled by a
``background model'' $\psi$, a broad nonspecific distribution covering
overall gene-set activity across the whole dataset repository.  All
weights are non-negative and $\sum_{j=1}^{N_S + 1} \theta^q_j =
1$. In essence, this representation assumes that biological activity
in the query dataset can be approximately explained as a combination
of earlier datasets and a novelty term.

The remaining task is to infer the combination model
$\boldsymbol{\Theta}^q$ for each query $q$ given the known models
$M^{s_j}$ of datasets in the repository.  We infer a maximum a
posteriori (MAP) estimate of the weights
$\boldsymbol{\Theta}^q=\{\theta^q_j\}_{j=1}^{N_S + 1}$.  Alternatively
we could sample over the posterior, but MAP inference already yielded
good results.  We optimize the combination weights to maximize their
(log) posterior probability
\begin{align} \label{costfun}
&\mbox{log }p(\{\theta^q_j\}|\{x^q_i\}, \{M^{s_j}\})
\propto \mbox{log }p(\{x^q_i\}|\{M^{s_j}\},\{\theta^q_j\}) 
+ \mbox{log }p(\{\theta^q_j\}) \nonumber\\
&\propto \sum_i \mbox{log} \Big[\Big(\sum_{j=1}^{N_S} \theta^q_j
    p(x_i^q | M^{s_j}) \Big) + \theta^q_{N_S + 1} p(x_i^q|\psi)
\Big] 
- \lambda\sum_{j=1}^{N_S + 1} {\theta^q_j}^2
\end{align}
where $p(\{\theta^q_j\}) = \mathcal{N}(0,\lambda^{-1}\boldsymbol{I})$ is 
a naturally non-sparse $L_2$ prior for the weights with a regularization
term $\lambda$.  
The cost function \eqref{costfun} is strictly concave (\emph{SI
  Text}), and standard constrained convex optimization techniques can
be used to find the optimized weights. Algorithmic details for the
Frank-Wolfe algorithm and a proof of convergence are provided in
\emph{SI Text}.  After computing the MAP estimate, we rank the
datasets for retrieval according to decreasing combination weights.

This modeling-driven approach has several advantages: 1) the
approximations become more accurate as more datasets are submitted to
the repository, increasing naturally the number of base distributions;
2) it is fast, since only the models of the datasets are needed, not
the large datasets themselves; 3) any model types can be included, as
long as likelihoods of an observed sample can be computed, hence all
expert knowledge built into the models in the repository can be used;
4) relevant datasets are not assumed to be ``similar'' to the query in
any na\"ive sense, they only need to explain a part of the query set;
5) the relevance scores of datasets have a natural quantitative
meaning as weights in the probabilistic combination model.

\subsection{Scalability} 
As the size of repositories such as ArrayExpress doubles every two
years or even faster \cite{Parkinson09}, fast computation with respect
to the number $N_S$ of background datasets is crucial for future-proof
search methods.  Already the first method above has a fast linear
computation time in $N_S$ (\emph{SI Text}), and an approximate variant
can be run in sublinear time. For that, the model combination will be
optimized only over the $k$ background datasets most similar to the
query, which can be found in time $O(N_S^{1/(1+\epsilon)})$ where
$\epsilon\ge 0$ is an approximation parameter \cite{Gionis99}, by
suitable hashing functions.

\section{Results}
\subsection{Data-driven retrieval of experiments is more accurate than
  standard keyword search}

\begin{figure}
\centering
\includegraphics{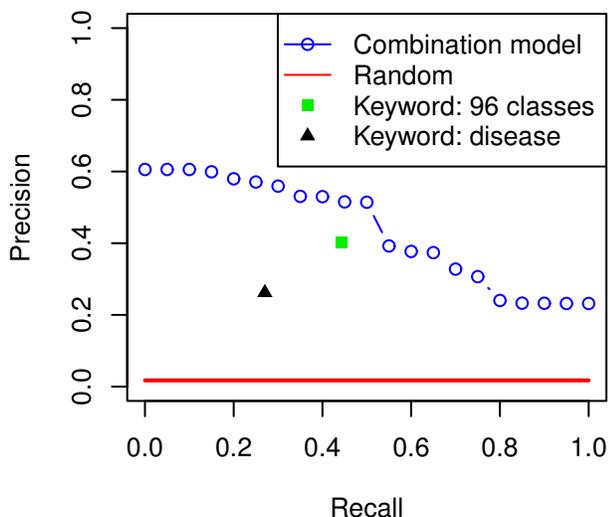}
\caption{Data-driven retrieval outperforms
  the state of the art of keyword search on the human gene expression
  atlas \cite{Lukk12}. Blue: Traditional precision-recall curve where
  progressively more datasets are retrieved from left to right. All
  experiments sharing one or more of the 96 biological categories of
  the atlas were considered relevant. In keyword retrieval, either the
  category names (``Keyword: 96 classes'') or the disease annotations
  (``Keyword: disease'') were used as keywords. All datasets having at
  least ten samples were used as query datasets, and the curves are
  averages over all queries.
\label{fig:PR_nonsmallVsall}}
\end{figure}

We benchmarked the combination model against state-of-the-art dataset
retrieval by keyword search, in the scenario where a user queries with
a new dataset against a database of earlier released datasets
represented by models. The data were from a large human gene
expression atlas \cite{Lukk12}, containing 206 public datasets with
$5372$ samples in total that have been systematically annotated and
consistently normalized. To make use of prior biological knowledge, we
preprocessed the data by gene set enrichment analysis
\cite{Subramanian05}, representing each sample by an integer vector
telling for each gene set the number of leading edge active genes
\cite{Caldas09} (\emph{Methods}). As base models we used two model types previously
applied in gene expression analysis \cite{Gerber07, Caldas09,
  Engreitz10, Caldas12}: a discrete principal component analysis
method called Latent Dirichlet Allocation \cite{Pritchard00, Blei03},
and a simpler variant called mixture of unigrams \cite{Nigam00}
(\emph{SI Text}). Of the two types, for each dataset we
chose the model yielding the larger predictive likelihood
(\emph{SI Text}).  For each query ($q$), the earlier
datasets ($s_j$) were ranked in descending order of the combination
proportion ($\theta_j^q$; estimated from Eq. \eqref{costfun}). That
is, base models which explained a larger proportion of the gene set
activity in the query were ranked higher.  The approach yields good
retrieval: the retrieval result was consistently better than with
keyword searches applied to the titles and textual descriptions of the
datasets (Fig.~\ref{fig:PR_nonsmallVsall}), which is a standard
approach for dataset retrieval from repositories
\cite{Zhu08geometadb}.

We checked that the result is not only due to laboratory effects by
discarding, in a follow-up study, all retrieved results from the same
laboratory. The mean average precision decreased slightly (from $0.44$
to $0.42$; precision-recall curve in Fig.~S2) but still
supports the same conclusion.

\subsection{Network of computationally recommended dataset connections
  reveals biological relationships}

\begin{figure*}
\centering
\includegraphics[angle=270]{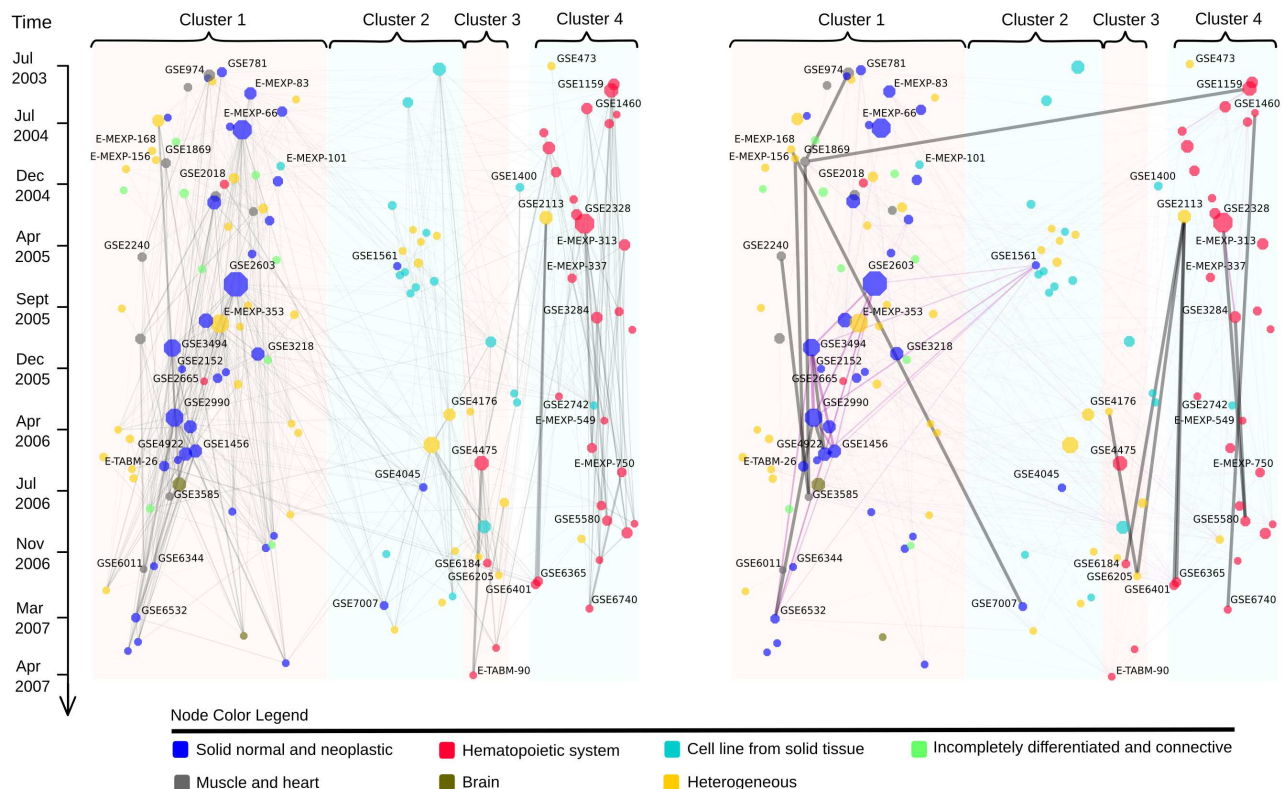}
\caption{Relevance network of datasets in the human gene expression
  atlas; data-driven links from the model (left) and citation links
  (right). Left: each dataset was used as a query to retrieve earlier
  datasets; a link from an earlier dataset to a later one means 
  the earlier dataset is relevant as a partial model of activity in 
  the later dataset. Link width is proportional to the normalized relevance
  weight (combination weight $\theta_{j}^q$; only links with
  $\theta_{j}^q\geq 0.025$ are shown, and datasets without links have
  been discarded). Right: links are direct (gray) and indirect
  (purple) citations.  Node size is proportional to the estimated
  influence, \emph{i.e.}, the total outgoing weight. Colors: tissue
  types (six meta tissue types \cite{Lukk12}). The node layout was
  computed from the data-driven network (details in
  \emph{Methods}).\label{fig:full_network}}
\end{figure*}

When each dataset in turn is used as a query, the estimated
combination weights form a ``relevance network'' between datasets
(Fig.~\ref{fig:full_network} left), where each dataset is linked to
the relevant earlier datasets (for details, see {\it Methods}; a
larger figure in Fig.~S5; and an interactive searchable version in the
online resource). The network structure is dominated but not fully
explained by the tissue type.  Normal and neoplastic solid tissues
(cluster 1) are clearly separate from cell lines (cluster 2) and from
hematopoietic tissue (cluster 4); the same main clusters were observed in
\cite{Lukk12}. Note that the model has not seen the tissue types but
has found them from the data. In closer inspection of the clusters,
some finer structure is evident. The muscle and heart datasets (gray)
form an interconnected subnetwork in the left edge of the image: nodes
near the bottom of the image (downstream) are explained by earlier
(upstream) nodes, which in turn are explained by nodes even further
upstream. As another example, in cluster 4 myeloma and leukemia
datasets are concentrated on the left side of the cluster, whereas the
right side mostly contains normal or infected mononuclear cells.

There is a substantial number of links both across clusters and across
tissue categories.  Among the top thirty cross-category links, 25
involve heterogeneous datasets containing samples from diverse tissue
origins.  The strongest link connects GSE6365, a study on multile
myeloma, with GSE2113, a larger study from the same lab which largely
includes the GSE6365 samples. The dataset E-MEXP-66 is a hub connected
to all the clusters and to nodes in its own cluster having different
tissue labels.  It contains samples studying Kaposi sarcoma, and
includes control samples from skin endothelial cells from blood
vessels and the lymph system. Blood vessels and cells belonging to the
lymph system are expected to be present in almost any solid tissue
biopsy as well as in samples based on blood samples.  The strongest
link between two homogeneous datasets of different tissue types
connects GSE3307 (which compares skeletal muscle samples from healthy
individuals with 12 groups of patients affected by various muscle
diseases) to GSE5392, which measures transcriptome profiles of normal
brain and brain bipolar disorder. Interestingly, shortening of
telomeres has been associated both with bipolar disorder
\cite{Martinsson13} and muscular disorder
\cite{Mourkioti13}. Treatment of bipolar disorder has been found to
also slow down the onset of skeletal muscle disorder
\cite{Kitazawa08}.
 
Next we investigated ``outlier" datasets where the tissue type does
not match the main tissue types of a cluster, implying that they might
reveal commonalities between cellular conditions across tissues.
Cluster~1 contained three outlier datasets: two hematopoietic datasets
and one cell line dataset.  The two hematopoietic outlier datasets are
studies related to macrophages and are both strongly connected to
GSE2004, which contains samples from kidney, liver, and spleen, sites
of long-lived macrophages.  The first hematopoietic outlier, GSE2018
studies bronchoalveolar lavage cells from lung transplant receipts;
the majority of these cells are macrophages.  The dataset has strong
links to solid tissue datasets including
GSE2004, and the diverse dataset E-MEXP-66. The second
hematopoietic outlier, GSE2665, is also strongly connected to GSE2004
and measures expression of lymphatic organs (sentinel lymph node) that
contain sinusoidal macrophages and sinusoidal endothelial cells.  The
third outlier, E-MEXP-101, studies a colon carcinoma cell line and has
connections to other cancer datasets in cluster~1.

\subsection{Top dataset links overlap well with citation graph} 

We compared the model-driven network to the actual citation links
(Fig.~\ref{fig:full_network}, right) to find out to what extent the
citation practice in the research community matches the data-driven
relationships. Of the top two hundred data-driven edges,  50\% overlapped
with direct or indirect citation links (see \emph{Methods} and
\emph{SI Text}).  Most of the direct citations appear
within the four tissue clusters (Fig.~\ref{fig:full_network},
right). The two cross-cluster citations are not due to biological
similarity of the datasets. The publication for GSE1869 cites the
publication for GSE1159 regarding the method of differential
expression detection. The GSE7007, a study on Ewing sarcoma samples, cites
the study on human mesenchymal stem cells (E-MEXP-168) for stating
that the overall gene expression profiles differ between those samples.

We additionally compared the densely connected sets of experiments 
between the two networks. 
In the citation graph the breast cancer datasets GSE2603, GSE3494, GSE2990, 
GSE4922, and GSE1456 form an interconnected clique in cluster 1, while 
the three leukocyte datasets GSE2328, GSE3284, and GSE5580 form an 
interconnected module in cluster 4. In the relevance network the corresponding 
edges for both cliques are among the strongest links for those datasets,
and some of them are among the top 20 strongest edges in the network 
(see \emph{SI Text} for the list of top 20 edges). 
There are also densely connected modules in the relevance network that are 
not strongly connected in the citation graph; when we systematically sought 
cliques associated to each of the top 20 edges, the most strong edges 
constitute a clique among E-MEXP-750, GSE6740 and GSE473, all three studying CD4+ T 
helper cells which are an essential part of the human immune system. 
Another  
interesting set is among three T-cell related datasets in cluster 3. Two of the 
datasets contain T lymphoblastic leukemia samples (E-MEXP-313 and E-MEXP-549), whereas 
E-MEXP-337 reports thymocyte profiles. Thymocytes are developing T lymphocytes that 
are matured in thymus, so this connection is biologically meaningful but not
straightforward to find from dataset annotations. Other strongly connected cliques are 
discussed in the \emph{SI Text}.

\subsection{Analysis of network hubs discovers datasets deserving more
  citations}

\begin{figure}
\centering
\includegraphics[angle=270]{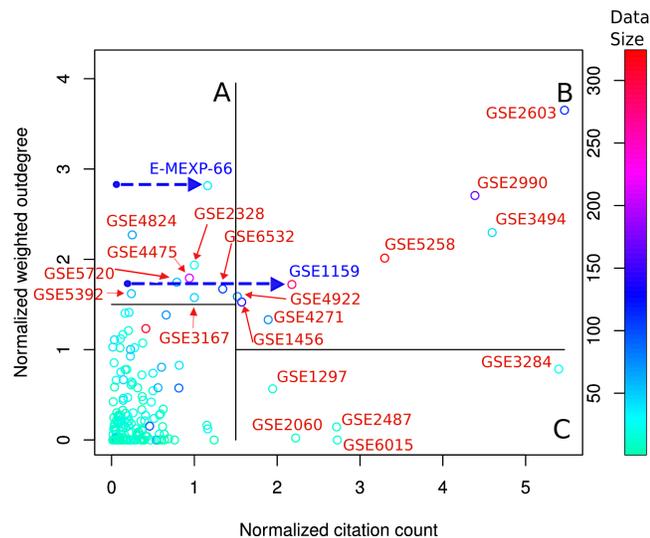}
\caption{Data-driven prediction of usefulness of datasets vs. their
  citation counts.  Manual checks comparing sets for which the two
  scores differed revealed inconsistent database records for two
  datasets; the blue arrows point to their corrected locations, which
  are more in line with the data-driven model. Regions A, B, and C: see
  text.\label{fig:normalized_scatter_plot}}
\end{figure}

Datasets that have high weights in explaining other datasets have a
large weighted outdegree in the data-driven relevance network, and are
expected to be useful for many other studies. We checked whether the
publications corresponding to these \emph{central hubs} are highly
cited in the research community. There is a low but statistically
significant correlation between the weighted outdegree of datasets and
their citation counts (Fig.~\ref{fig:normalized_scatter_plot};
Spearman $\rho(169) = 0.2656$, $p < 0.001$).  Both quantities were
normalized to avoid bias due to different release times of the
datasets (\emph{Methods}).  We further examined whether the prestige
of the publication venue (measured by impact factor) and the senior
author (h-index of the last author) biased the citation counts, which
could explain the low correlation between the outdegree and the
citation count, and the answer was affirmative (\emph{Methods}).

We inspected more closely the datasets where the recommended or the
actual citation counts were high
(Fig.~\ref{fig:normalized_scatter_plot}): (A) datasets having low
citation counts but high outdegrees, (B) both high citation counts and
high outdegrees and (C) high citation counts but low outdegrees.  We
manually checked the publication records of region A in Gene Expression
Omnibus (GEO) \cite{Barrett11} and ArrayExpress \cite{Parkinson09}, to
find out why the datasets had low citation counts despite their high
outdegree (data-driven citation recommendations).  Two of the eight
datasets had an inconsistent publication record. The blue arrows in
Fig.~\ref{fig:normalized_scatter_plot} point from their original
position to the corrected position confirmed by GEO and ArrayExpress.
Thus the data-driven network revealed the inconsistency, and the new
positions, corresponding to higher citation counts, validate the
model-based finding that these datasets are good explainers for other
datasets.  In region B, most of the papers have been published in high
impact journals and have relatively high number of samples (average
sample size of $154$) compared to region A (average sample size of
$75$). One of the eight datasets in the collection is the well known
Connectivity Map experiment (GSE5258).  Lastly the set C mostly contains unique
targeted studies; there are five studies in the set, which are about
leukocytes of injured patients, Polycomb group (PcG) proteins,
senescence, Alzheimer's disease, and effect of cAMP agonist forskolin,
a traditional Indian medicine. The studies have been published in high
impact forums, and a possible reason of their low outdegree is their
specific cellular responses, which are not very common in the atlas.

\section{Discussion}

Our main goal was to test the feasibility of the scenario where
researchers let the data speak for themselves when relating new
research to earlier studies.  The conclusion is positive: even a
relatively straightforward and scalable mixture modeling approach
found both expected relationships such as tissue types, and
relationships not easily found with keyword searches, including cells
in different developmental stages or treatments resembling conditions
in other cell types.  While biologists could find such connections by
bringing expert knowledge into keyword searches, the ultimate
advantage of the data-driven approach is that it also yields
connections beyond current knowledge, giving rise to new hypotheses
and follow-up studies. For example, it seems surprising that the
skeletal muscle dataset GSE6011 is linked also to 
kidney and brain datasets. 
Closer inspection yielded possible partial
explanations.  Some kidney areas are rich in blood vessels, lined by
smooth muscle.  Studies have shown common gene signatures between
skeletal muscle and brain.  Abnormal expression of the protein
dystrophin leads to Duchenne muscular dystrophy, exhibited by a
majority of samples in GSE6011; the brain is another major expression
site for dystrophin \cite{Culligan2001}.  
Interestingly the top three potentially novel datasets, where only less than 50\% of 
the expression pattern is modelled by earlier datasets (i.e. 
$\theta_{N_S + 1}^q > 0.5$), are GSE2603 (a central breast cancer set), 
the Connectivity Map data (GSE5258) and the Burkitt's Lymphoma set 
(GSE4475; a cancer fundamentally distinct from other types of lymphoma). 
The first two are also recovered by the citation data (have 
relatively high citation counts and appear in region B in
Fig.~\ref{fig:normalized_scatter_plot}), unlike the third (which is
part of region A in Fig.~\ref{fig:normalized_scatter_plot}).

Our case study focused on global analysis of the relevance network
obtained for a representative dataset collection, allowing for
comparisons with the citation graph.  The data-driven relationships
corresponded to actual citations when available, but were richer and
were able to spot out errors in citation links. Another intended use
of the retrieval method is to support researchers in finding relevant
data on a particular topic of interest. We performed a study to obtain
insights into relationships among skeletal muscle datasets as well as
between skeletal muscle and other datasets, and showed that the
retrieval method lessens the need for laborious manual searches
(\emph{SI Text} and Fig.~S4).

In this work we made simplifying assumptions: we only employed two model
families, included biological knowledge only as pre-chosen gene sets,
and assumed all new experiments to be mixtures of earlier ones,
instead of sums of effects in them. We expect results to improve
considerably with more advanced future alternatives, with the research
challenge being to maintain scalability.  Generalizability of the
search across measurement batches, laboratories, and measurement
platforms is a challenge.  Our feasibility study showed that for
carefully preprocessed datasets (of the microarray
atlas~\cite{Lukk12}), data-driven retrieval is useful even across
laboratories.  Our method is generally applicable to any single
platform, and takes into account the expert knowledge built into
models of datasets for that platform; abstraction-based data
representations, such as the gene set enrichment representation we
used, have potential to facilitate cross-platform analysis.  As data
integration approaches develop
further~\cite{Tripathi11dami,Virtanen12aistats}, it may be possible to
do searches even across different omics types; here, integration of
meta data (pioneered in a specific semi-supervised framework
\cite{Wise2012}), several ontologies (MGED ontology, experimental
factor ontology and ontology of biomedical investigations
\cite{Zheng11}) and text mining results \cite{Jensen06, Rzhetsky08}
are obviously useful first steps. 

\begin{materials}
\section{Gene expression data}
We used the human gene expression atlas \cite{Lukk12} available at
ArrayExpress under accession number E-MTAB-62. The
data were preprocessed by gene set enrichment analysis (GSEA) using the
canonical pathway collection (C2-CP) from the Molecular Signatures
Database \cite{Subramanian05}. Each sample was represented by its 
top enriched gene sets \cite{Caldas09} (\emph{SI Text}).

\section{Node layout and normalized relevance weight}
The weight matrix contains a weight vector for each query dataset,
encoding the amount of variation in that query explained by each
earlier dataset. As query datasets from early years have only few even
earlier sets available, there is a bias towards the edges being
stronger for the datasets from early years. To remove the bias we
normalized, for the visualizations, the edge strengths of each query
data set by the number of earlier datasets.  To visualize the
relationship network over time in Fig.~\ref{fig:full_network}, we
needed a layout algorithm that positions the datasets on the
horizontal axis highlighting structure and avoiding tangling. We used
a \emph{cluster-emphasizing} Sammon's mapping \cite{Sammon69};
Sammon's mapping is a nonlinear projection method or Multidimensional
Scaling algorithm which aims at preserving the interpoint distances
(here $1-\theta_j^q$). By clustering the network (with unsupervised
Markov clustering \cite{Dongen00}) and increasing between-cluster
distances by adding a constant ($c=1$) to them, the mapping was made
to emphasize clusters and hence untangle the layout.

\section{Citation graph} 
Direct citations between dataset-linked publications were extracted
from the Web of Science (26 Jul 2012) and PubMed (17 Oct 2012).  We
additionally considered two types of indirect edges.  Firstly, we
introduced links between datasets whose publications share common
references.  This covers for instance related datasets whose
publications appeared close in time, making direct citation unlikely.
A natural measure of edge strength is given by the number of shared
references.  Secondly, we connect datasets whose articles are cited
together, because co-citation is a sign that the community perceives
the articles as related. Here, the edge strength was taken to be the
number of articles co-citing the two dataset publications; these edges
dominate the indirect links in the citation graph.  For this analysis
we used citation data, available for $171$ datasets and provided by 
Thomson Reuters as of 13 September 2012.

\section{Normalization of citation counts and weighted outdegrees}
As early datasets have many more papers which can cite them, and many
more later datasets which they can help model, both the citation
counts and estimated weighted outdegrees are expected to be upwards
biased for them. For Fig.~\ref{fig:normalized_scatter_plot} we
normalized the quantities; for each dataset we normalized the
outdegree by the number of newer datasets, and the citation count by
the time difference between publishing the data and the newest dataset
in the atlas. To make sure the normalization did not introduce
side effects we additionally checked that the same conclusions were
reached 
without the citation count normalization (Fig.~S1; plotted 
as stratified subfigures for each 1-year time window).
The citation
counts were extracted from PubMed on 16 May 2012.

\section{Citation counts are strongly influenced by external esteem of
the publication forum and the senior author} We stratified the data
sets according to the numbers of data-driven citation recommendations,
and studied whether the impact factor of the forum or the h-index of
the last author were predictive of the actual citation count in each
stratum. The strata were the top and bottom quartiles, and for each we
compared the top and bottom quartiles of the actual citation counts
(resulting in comparing the four corners of
Fig.~\ref{fig:normalized_scatter_plot}). 
For low outdegree (low recommended citation count), 
the h-index was lower for less cited datasets
($t_{11} = 2.78, p = 0.0086$; mean value $24.20$ vs $54.62$), and also
the impact factor was lower ($t_{7} = 2.6, p = 0.016$; mean value
$4.38$ vs $21.13$). Similarly, for high recommended citation count 
the impact factor for the little-cited datasets was 
lower ($t_{19} = 3.99, p = 4.0^{-4}$; mean
value $6.45$ vs $21.91$), while the difference in h-index was not
significant. All t statistics and p-values were computed by one-sided
independent sample Welch's t-tests. The h-indices and impact factors
were collected from Thomson Reuters Web of Knowledge and Journal
Citation Reports 2011 respectively on 23rd July 2012.

\end{materials}

\begin{acknowledgments}
 We thank Matti Nelimarkka and Tuukka Ruotsalo for helping with
 citation data.  Certain data included herein are derived from the
 following indices: Science Citation Index Expanded, Social Science
 Citation Index and Arts \& Humanities Citation Index, prepared by
 Thomson Reuters\textsuperscript{\textregistered}, Philadelphia,
 Pennsylvania, USA, \textsuperscript{\copyright} Copyright Thomson
 Reuters \textsuperscript{\textregistered}, 2011.  This work was
 financially supported by the Academy of Finland (Finnish Centre of
 Excellence in Computational Inference Research COIN, grant no
 251170).
\end{acknowledgments}

\end{article}


\begin{thebibliography}{10}

\bibitem{Greene11}
Greene C-S, Troyanskaya O-G (2011) 
\newblock {PILGRM}: An interactive data-driven discovery platform for expert
  biologists.
\newblock {\em Nucleic Acids Res} 39:W368--74.

\bibitem{Tanay05}
Tanay A, Steinfeld I, Kupiec M, Shamir R (2005) 
\newblock Integrative analysis of genome-wide experiments in the context of a
  large high-throughput data compendium.
\newblock {\em Mol Syst Biol} 1:e1--10.

\bibitem{Caldas12}
Caldas~J, {et al.} (2012) 
\newblock Data-driven information retrieval in heterogeneous collections of
  transcriptomics data links \emph{SIM2s} to malignant pleural mesothelioma.
\newblock {\em Bioinformatics} 28:i246--i253.

\bibitem{Adler09}
Adler~P, {et al.} (2009) 
\newblock Mining for coexpression across hundreds of datasets using novel rank
  aggregation and visualization methods.
\newblock {\em Genome Biol} 10:R139.

\bibitem{Schmid12}
Schmid P-R, Palmer N-P, Kohane I-S, Berger B (2012) 
\newblock Making sense out of massive data by going beyond differential
  expression.
\newblock {\em Proc Natl Acad Sci U S A} 109:5594--5599.

\bibitem{Gerber07}
Gerber G-K, Dowell R-D, Jaakkola T-S, Gifford D-K (2007) 
\newblock Automated discovery of functional generality of human gene expression
  programs.
\newblock {\em PLoS Comput. Biol} 3:e148.

\bibitem{Tseng12}
Tseng G-C, Ghosh D, Feingold E (2012) 
\newblock Comprehensive literature review and statistical considerations for
  microarray meta-analysis.
\newblock {\em Nucleic Acids Res} 40:3785--3799.

\bibitem{Rung12}
Rung J, Brazma A (2012) 
\newblock Reuse of public genome-wide gene expression data.
\newblock {\em Nat Rev Genet} 14:89--99.

\bibitem{Baxter97}
Baxter J (1997) 
\newblock A {B}ayesian/information theoretic model of learning to learn via
  multiple task sampling.
\newblock {\em Machine Learning} 28:7--39.

\bibitem{Caruana97}
Caruana R (1997) 
\newblock Multitask learning.
\newblock {\em Machine Learning} 28:41--75.

\bibitem{Punta12}
Punta~M, {et al.} (2012) 
\newblock The {P}fam protein families database.
\newblock {\em Nucleic Acids Res} 40:D290--D301.

\bibitem{Lukk12}
Lukk~M, {et al.} (2010) 
\newblock A global map of human gene expression.
\newblock {\em Nat Biotechnol} 28:322--324.

\bibitem{Russ10}
Russ J, Futschik M-E (2010) 
\newblock Comparison and consolidation of microarray data sets of human tissue
  expression.
\newblock {\em BMC Genomics} 11:305.

\bibitem{plosDisease}
Suthram~S, {et al.} (2010) 
\newblock Network-based elucidation of human disease similarities reveals
  common functional modules enriched for pluripotent drug targets.
\newblock {\em PLoS Comput Biol} 6:e1000662.

\bibitem{Huttenhower08}
Huttenhower C, Troyanskaya O-G (2008) 
\newblock Assessing the functional structure of genomic data.
\newblock {\em Bioinformatics} 24:i330--8.

\bibitem{Meinicke11}
Meinicke P, Asshauer K-P, Lingner T (2011) 
\newblock Mixture models for analysis of the taxonomic composition of
  metagenomes.
\newblock {\em Bioinformatics} 27:1618--24.

\bibitem{Parkinson09}
Parkinson H, {et al.} (2009)
\newblock ArrayExpress update--from an archive of functional genomics
experiments to the atlas of gene expression.
\newblock {\em Nucleic Acids Res} 37:D868--72.

\bibitem{Gionis99}
Gionis A, Indyk P, Motwani R (1999) 
\newblock Similarity search in high dimensions via hashing.
\newblock {\em Proceedings of the Twenty-Fifth International Conference on Very Large Databases} 
eds Atkinson M, Orlowwska M, Valduriez P, Zdonik S, Brodie M (Edinburgh, Scotland) pp 518--529. 

\bibitem{Subramanian05}
Subramanian~A, {et al.} (2005) 
\newblock Gene set enrichment analysis: {A} knowledge-based approach for
  interpreting genome-wide expression profiles.
\newblock {\em Proc Natl Acad Sci U S A} 102:15545--15550.

\bibitem{Caldas09}
Caldas J, Gehlenborg N, Faisal A, Brazma A, Kaski S (2009) 
\newblock Probabilistic retrieval and visualization of biologically relevant
  microarray experiments.
\newblock {\em Bioinformatics} 25:i145--i153.

\bibitem{Engreitz10}
Engreitz J-M, {et al.} (2010) 
\newblock Content-based microarray search using differential expression
  profiles.
\newblock {\em BMC Bioinformatics} 11:603.

\bibitem{Pritchard00}
Pritchard J-K, Stephens M, Donnelly P (2000) 
\newblock Inference of population structure using multilocus genotype data.
\newblock {\em Genetics} 155:945--959.

\bibitem{Blei03}
Blei D-M, Ng~A-Y, Jordan M-I, Lafferty J (2003) 
\newblock {L}atent {D}irichlet allocation.
\newblock {\em J Mach Learn Res} 3:993--1022.

\bibitem{Nigam00}
Nigam K, McCallum A, Thrun S, Mitchell T (2000) 
\newblock Text classification from labeled and unlabeled documents using {EM}.
\newblock {\em Machine Learning} 39:103--134.

\bibitem{Zhu08geometadb}
Zhu Y, Davis S, Stephens R, Meltzer P-S, Chen Y (2008) 
\newblock {GEO}metadb: powerful alternative search engine for the {G}ene
  {E}xpression {O}mnibus.
\newblock {\em Bioinformatics} 24:2798--2800.

\bibitem{Mourkioti13}
Mourkioti F, {et al.} (2013)
\newblock Role of telomere dysfunction in cardiac failure in {D}uchenne muscular dystrophy.
\newblock {\em Nature Cell Bio} 15:895--904.

\bibitem{Kitazawa08}
Kitazawa M, Trinh D-N, LaFerla F-M (2008)
\newblock Inflammation induces tau pathology in inclusion body myositis model via glycogen synthase kinase-3 beta.
\newblock {\em Ann Neurol} 64:15--24.

\bibitem{Martinsson13}
Martinsson L, {et al.} (2013)
\newblock Long-term lithium treatment in bipolar disorder is associated with longer leukocyte telomeres.
\newblock {\em Transl Psychiatry} 3:e261.  

\bibitem{Kimura1995}
Kimura~K, {et al.} (1995) 
\newblock Diversity and variability of smooth muscle phenotypes of renal
  arterioles as revealed by myosin isoform expression.
\newblock {\em Kidney Int} 48:372--382.

\bibitem{Culligan2001}
Culligan K, Glover L, Dowling P, Ohlendieck K (2001) 
\newblock Brain dystrophin-glycoprotein complex: Persistent expression of
  beta-dystroglycan, impaired oligomerization of {Dp71} and up-regulation of
  utrophins in animal models of muscular dystrophy.
\newblock {\em BMC Cell Biol} 2:2.

\bibitem{Kirchner1988}
Kirchner~T, {et al.} (1988) 
\newblock Pathogenesis of myasthenia gravis. acetylcholine receptor-related
  antigenic determinants in tumor-free thymuses and thymic epithelial tumors.
\newblock {\em Am J Pathol} 130:268--280.

\bibitem{Holliday2011}
Holliday D-L, Speirs V (2011) 
\newblock Choosing the right cell line for breast cancer research.
\newblock {\em Breast Cancer Res} 13:215.

\bibitem{Barrett11}
Barrett~T, {et al.} (2011) 
\newblock {NCBI GEO: archive for functional genomics data sets-10 years on.}
\newblock {\em Nucleic Acids Res} 39:D1005--D1010.

\bibitem{Tripathi11dami}
Tripathi A, Klami A, Ore{\v{s}}i{\v{c}} M, Kaski S (2011) 
\newblock Matching samples of multiple views.
\newblock {\em Data Min Knowl Discov} 23:300--321.

\bibitem{Virtanen12aistats}
Virtanen S, Klami A, Khan S-A, Kaski S (2012) 
\newblock Bayesian group factor analysis.
\newblock {\em JMLR Workshop Conf Proc} 22:1269--1277.

\bibitem{Wise2012}
Wise A, Oltvai Z, Bar-Joseph Z (2012) 
\newblock Matching experiments across species using expression values and
  textual information.
\newblock {\em Bioinformatics} 28:i258--i264.

\bibitem{Zheng11}
Zheng~J, {et al.} (2011) 
\newblock Annotcompute: annotation-based exploration and meta-analysis of
  genomics experiments.
\newblock {\em Database (Oxford)}.

\bibitem{Jensen06}
Jensen L-J, Saric J, Bork P (2006) 
\newblock Literature mining for the biologist: from information retrieval to
  biological discovery.
\newblock {\em Nat Rev Genet} 7:119--129.

\bibitem{Rzhetsky08}
Rzhetsky A, Seringhaus M, Gerstein M (2008) 
\newblock Seeking a new biology through text mining.
\newblock {\em Cell} 134:9--13.

\bibitem{Sammon69}
Sammon J-W (1969) 
\newblock A nonlinear mapping for data structure analysis.
\newblock {\em IEEE Trans Comput} 18:401--409.

\bibitem{Dongen00}
van Dongen~S (2000) 
\newblock Graph Clustering by Flow Simulation.
\newblock {\em PhD thesis, University of Utrecht}.

\end{thebibliography}
\end{document}


\title{Supporting Information Appendix: Toward computational cumulative biology 
by combining models of biological datasets}

\author{Ali Faisal\affil{1}{Helsinki Institute for Information Technology HIIT, Department of Information and Computer Science, PO Box 15400, Aalto University, 00076 Aalto, Finland}, Jaakko Peltonen\affil{1}{}, Elisabeth Georgii\affil{1}{}, Johan Rung\affil{2}{European Molecular Biology Laboratory, European Bioinformatics Institute (EMBL-EBI), Wellcome Trust Genome Campus Hinxton, CB10 1SD, UK} \and Samuel Kaski\affil{1}{}\affil{3}{Helsinki Institute for Information Technology HIIT, Department of Computer Science, PO Box 68, 00014 University of Helsinki, Finland}}
\maketitle

\begin{article}
\section{Methods}
\paragraph{Gene Set Enrichment Analysis.}  We used GSEA
\cite{Subramanian05} to bring in biological knowledge in the form of
pre-defined gene sets.  GSEA starts by sorting genes with
respect to their normalized expression levels.  GSEA essentially
consists of computing a running sum on the sorted list for each gene
set; this running sum (enrichment score) increases when a gene belongs
to the gene set and decreases otherwise; the final statistic is the
maximum of this running sum. The procedure essentially amounts to
computing a weighted Kolmogorov-Smirnov (KS) statistic.  For each
sample the KS statistic is normalized by dividing it by the mean of
random KS statistics computed on randomly generated gene sets whose
size is matched with the actual gene set; the $50$ top-scoring gene
sets are selected according to this normalized score.  This simple
thresholding ignores significance values but has been successfully
used in earlier meta-analysis studies \cite{Segal04, Caldas09,
  Caldas12}; we earlier investigated the alternative of selecting gene
sets based on a standard q-value cut-off (with $q < 0.05$), but it
produced an excessively sparse encoding where more than $80\%$
samples had no active gene sets \cite{Caldas12}. The activity of
each gene set is finally expressed as its core set or \emph{leading
  edge subset}, consisting of genes found before the running KS score
reached its maximum. We quantify the activity or a set simply by the
size of the leading edge subset. It can be used analogously to the so
called \emph{bag-of-words} representations in text analysis, and we
use it for the subsequent modeling of each dataset with the base models.

\paragraph{Base models.}  \emph{Latent Dirichlet allocation} (LDA)
\cite{Pritchard00, Blei03, Griffiths04} and \emph{mixture of unigrams}
\cite{Nigam00} are probabilistic unsupervised models that give insight
to datasets by describing them in terms of latent components.  Each
data sample, in our case quantified as gene set activities, is
represented by a probability distribution over components (sometimes
also called topics). The components are shared by all samples, but
with different degrees of activation for each, and each component
produces a characteristic distribution over the gene sets.  In LDA
each sample may be produced by multiple hidden components while the
mixture of unigrams is a simplified version where each sample is
assumed to come from a single component.  The computational problem is
to estimate the latent component structure (for each sample, the
distribution over components and for each component, its distribution
over gene sets) that has most likely generated the observed set of
samples.

We use standard inference solutions for the models: collapsed Gibbs
sampler for the LDA \cite{Griffiths04} and Expectation Maximization
for the mixture of unigrams \cite{Nigam00}. In LDA, the
hyperparameters, the prior probability of each component, were
optimized with Minka's stable fixed point iteration scheme
\cite{Minka00}, interleaved with the collapsed Gibbs sampling of the
other parameters. The number of Gibbs iterations was $2500$, found to
yield good performance also earlier \cite{Blei03,
  Griffiths04, Teh06}. For the mixture of unigrams model, the maximum a
posterior solution is estimated with the Expectation Maximization (EM)
algorithm, using Laplace smoothing for the prior probabilities of the
mixture components \cite{Nigam00}.
 
\paragraph{Model selection for base models.}  For each dataset we
estimated the two base models and selected the one that best models the
dataset.  We split the dataset into two parts where 90\% of the
dataset samples were used for training the two models and the remaining
10\% samples to compute the test-set predictive likelihood, a
measure of how well the model fits the data.  We repeated this procedure
in a 10-fold cross-validation setup for each dataset and each of the
two models.  The model that performed better on average across the 10
folds was chosen to represent the dataset. Most datasets (74 out of 112)
preferred the more expressive LDA model.

The number of components was selected with cross-validation for both
models. We again used $10$-fold cross-validation, separately for each
dataset, to choose the number of components that lead to best
predictive likelihoods on test samples from the same dataset. We
observed that the optimal number of total topics could roughly be
summarized by $\lceil N_D/3 \rceil$ for LDA and $\lfloor \sqrt{N_D}
\rfloor$ for mixture of unigrams, where $N_D$ is the total number of
samples in a dataset.  All predictive likelihoods (both for model
selection and for retrieval) were computed using an empirical
likelihood method (see \cite{Li06}). For very small datasets
($N_D<10$) within-set cross-validation is very noisy and we chose LDA
which had been selected for the majority of datasets with $10 \leq
N_D \leq 15$.

\paragraph{Strict Concavity of the objective function.}
For each query dataset $q$, there are multiple samples
$i=1,\ldots,N_q$.  For each sample, each background dataset gives a
probability.  Let $\vec{x}_i =
[p(x_i^q|M^{s_1}),p(x_i^q|M^{s_2}),\ldots,p(x_i^q|M^{s_{N_S}}),p(x_i^q|\Psi)]^\top$
be the column vector of probabilities for sample $i$ from all
background datasets and from the novelty model, where $^\top$ denotes
vector transpose. Then the probability given by a mixture of
background datasets is $\vec{\theta}^\top \vec{x}_i$ where
$\vec{\theta}$ is the column vector of mixture weights (mixing
proportions) over the background datasets and the novelty model.
Since the mixing weights are mixture probabilities, they must lie in
the canonical simplex denoted as
\begin{equation}
\Delta = \left\{\boldsymbol{\theta} |
\theta_j \ge 0, \sum_{j=1}^{N_S+1} \theta_j = 1\right\} \;.
\end{equation}

Our optimization takes the form
\begin{multline}
\max_{\vec{\theta}\in \Delta} \log \left( \exp(-\lambda||\vec{\theta}||^2) \prod_i \vec{\theta}^\top \vec{x}_i \right) \\
=\max_{\vec{\theta}\in \Delta} \sum_i \log(\vec{\theta}^\top \vec{x}_i) -\lambda||\vec{\theta}||^2
=\max_{\vec{\theta}\in \Delta} f(\vec{\theta})
\end{multline}
where the objective function is
$$
f(\vec{\theta}) = \sum_i \log(\vec{\theta}^\top \vec{x}_i)-\lambda||\vec{\theta}||^2 \;.
$$ 

The objective $f(\vec{\theta})$ is a strictly concave function with respect to the 
multivariate parameter $\vec{\theta}$. This can be shown since the 
second derivative of the function is negative.
In detail, the function's gradient is
\begin{equation}
\nabla f(\vec{\theta}) = \sum_i \frac{\vec{x}_i}{\vec{\theta}^\top \vec{x}_i} - 2\lambda \vec{\theta} 
\end{equation}
and the matrix of second-order partial derivatives (Hessian matrix) is
\begin{equation}
\nabla^2 f(\vec{\theta}) = -\sum_i \frac{\vec{x}_i \vec{x}^\top_i}{(\vec{\theta}^\top \vec{x}_i)^2}
-2\lambda I
\end{equation}
where $I$ denotes the identity matrix. Since all elements of $\vec{x_i}$ are nonnegative,
it is easy to see the Hessian matrix is negative definite for all 
$\vec{\theta}$ 
where $\theta_i \ge 0$ 
as long as 
$\lambda>0$. Therefore the objective function is 
strictly concave.
A local maximum of a strictly concave function on a convex feasible region
is the unique global maximum \cite{Bradley77}; therefore maximizing the objective function
to a local maximum by any algorithm yields the unique global maximum.

\paragraph{Maximizing the Objective Function.}
Maximization of concave functions on the unit simplex $\Delta$ can be
done by the \emph{Frank-Wolfe algorithm}. The algorithm performs the
following steps.

\emph{Step 1: initialization.}  The algorithm initializes a solution
as the vertex of the simplex having the largest objective value. A
vertex of the simplex has $\theta_j=1$ for some $j$ and all other
elements of $\vec{\theta}$ are zero. At vertex
$j$, the objective function simplifies to
$$
\sum_i \log(x_{ij}) -\lambda
$$ where $x_{ij}=p(sample_i|dataset_j)$, which takes $O(N_q)$ time to
evaluate per vertex.  Thus creating the initialization takes linear
$O(N_q N_S)$ time even with a simple brute-force evaluation of all
vertices.

\emph{Step 2: Iteration.}  In each iteration, the algorithm improves
the solution by two steps: (1) Find the maximal element $j$ of the
gradient. (2) Find the point along the line $\vec{\theta} +
\alpha(\vec{e}(j)-\vec{\theta})$, $\alpha\in [0,1]$, which maximizes
the objective function. Here $\vec{e}(j)$ means the vector where the
element $j$ is one and the others are zero. Computation of the
gradient and finding the maximal element takes $O(N_q N_S)$ time.

\section{Proof of Convergence and Scalability}
\paragraph{Convergence Analysis.}
Since each iteration takes linear time with respect to the number of
query samples and background datasets, the only remaining issue is
the number of iterations required for good enough convergence. We now
analyze the convergence properties of this iteration in two cases.

\emph{Case 1: optimal $\alpha$.}  At first, we consider the case where
the best value of $\alpha$ can be found along the line to a sufficient
accuracy in a fixed amount of time, for example by restricting
evaluations along the line to a fixed number.

Define a proportional regret function $h(\vec{\theta})=(f(\vec{\theta}^{*})-f(\vec{\theta}))/4 C_f$ where $\vec{\theta}^{*}$ is the optimal parameter value maximizing the objective function,
and $C_f \ge 0$
is a \emph{measure of curvature} of $f$.
In detail, $C_f$ is defined as the largest quantity such
that for all $\vec{\theta}_A\in \Delta$, $\vec{\theta}_B\in \Delta$, $\vec{\theta}_C$ where $\vec{\theta}_C=\vec{\theta}_A + \alpha (\vec{\theta}_B-\vec{\theta}_A)$ for some $\alpha$, we have
$$
f(\vec{\theta}_C) \ge f(\vec{\theta}_A) + (\vec{\theta}_C-\vec{\theta}_A)^{\top} \nabla f(\vec{\theta}_A) - \alpha^2 C_f \;.
$$

With this notation, it can be shown (\cite{Clarkson12}, Theorem 2.2) that
at iteration $k$ of the Frank-Wolfe algorithm the current solution $\vec{\theta}_k$
has regret 
$$
h(\vec{\theta}_k) \le 1/(k+3)
$$
and thus
$$
f(\vec{\theta}^{*})-f(\vec{\theta}_k) \le 4C_f/(k+3) \;.
$$
Thus, to achieve a desired regret $\epsilon$, at most $4 C_f/\epsilon+3$
iterations are needed independently of the number of background datasets;
the amount of iterations needed depends only on the curvature.

\emph{Case 2: fixed $\alpha$.}
It can even be shown that using a fixed value
$\alpha_k=2/(k+3)$ at each iteration $k$ suffices to yield bounds for performance:
with this choice of $\alpha_k$ the regret bound at iteration $k+1$ becomes (\cite{Clarkson12}, Section 7)
$$
h(\vec{\theta}_{k+1}) \le 1/(k+4)\;
$$
and thus
$$
f(\vec{\theta}^{*})-f(\vec{\theta}_{k+1}) \le 4C_f/(k+4)
$$
which again shows the number of iterations to achieve a desired 
regret does not depend on the number of background datasets, 
only on the curvature. It is enough to show the curvature is finite and 
does not depend on the number of background datasets; we now show this.

\emph{Analysis of the Curvature: }
As seen above, the smaller the curvature $C_f$,
the better the bounds for the regret $f(\vec{\theta}^{*})-f(\vec{\theta}_{k})$ are.
In our case the function $f$ is twice differentiable and it can be shown
(\cite{Clarkson12}, Section 4.1) that
$$
C_f \le \sup_{\vec{\theta}_A\in \Delta,\vec{\theta}_B\in \Delta,\alpha\in[0,1]} -\frac{1}{2}(\vec{\theta}_B-\vec{\theta}_A)^\top \nabla^2 f(\vec{\theta}_\alpha) (\vec{\theta}_B-\vec{\theta}_A) \;.
$$
where $\vec{\theta}_\alpha = \vec{\theta}_A+\alpha(\vec{\theta}_B-\vec{\theta}_A)$.
For our cost function this becomes
\begin{multline}
C_f \le \sup_{\vec{\theta}_A\in \Delta,\vec{\theta}_B\in \Delta,\alpha\in[0,1]} \frac{1}{2}(\vec{\theta}_B-\vec{\theta}_A)^\top \\
\left(\sum_i \frac{\vec{x}_i \vec{x}^\top_i}{(\vec{\theta}_\alpha^\top \vec{x}_i)^2}
+2\lambda I
\right) (\vec{\theta}_B-\vec{\theta}_A) \\
= 
\sup_{\vec{\theta}_A\in \Delta,\vec{\theta}_B\in \Delta,\alpha\in[0,1]} \frac{1}{2} \\
\left( \sum_i \frac{((\vec{\theta}_B-\vec{\theta}_A)^\top\vec{x}_i)^2}{(\vec{\theta}_\alpha^\top \vec{x}_i)^2}
+2\lambda ||\vec{\theta}_B-\vec{\theta}_A||^2
\right) \\
\le 
\sup_{\vec{\theta}_A\in \Delta,\vec{\theta}_B\in \Delta,\alpha\in[0,1],1\le i \le N_q} \frac{1}{2} \bigg(
N_q \frac{((\vec{\theta}_B-\vec{\theta}_A)^\top\vec{x}_i)^2}{(\vec{\theta}_\alpha^\top \vec{x}_i)^2}
+4\lambda 
\bigg) \\
\le 
\sup_{\vec{\theta}_A\in \Delta,\vec{\theta}_B\in \Delta,\alpha\in[0,1],1\le i \le N_q} \frac{1}{2} \bigg(
N_q \frac{(\vec{\theta}_B\vec{x}_i)^2+(\vec{\theta}_A^\top\vec{x}_i)^2}{(\vec{\theta}_\alpha^\top \vec{x}_i)^2}
+4\lambda 
\bigg) \\
\le 
\sup_{\vec{\theta}_A\in \Delta,\vec{\theta}_B\in \Delta,\alpha\in[0,1],1\le i \le N_q} \frac{1}{2} \bigg(
N_q \frac{2 (\max_j x_{ij})^2}{(\min_j x_{ij})^2}
+4\lambda 
\bigg) 
\;
\end{multline}
where for brevity we denoted $x_{ij} = p(x_i^q|M^{s_j})$ for $j=1,\ldots,N_S$ 
and $x_{i,{N_S+1}}=p(x_i^q|\Psi)$.
Notice that the right-hand side only depends on the maximal and minimal values
that background datasets give to query samples, not on the number of such background 
datasets. Thus, as long as we ensure the maximal value is below some finite constant
and the minimal value is above some small nonzero constant, the curvature $C_f$ is finite 
and the convergence bounds of the Frank-Wolfe algorithm therefore do not depend on the number 
of background datasets. This condition is simple to ensure by suitable regularization 
of the models of background datasets.

Therefore under the simple condition that probabilities
given to query samples are upper bounded and lower bounded above zero,
the algorithm converges (towards the unique global maximum) to a
desired tolerance of the regret
in a finite number of iterations $I$, which can depend on the number of query samples but is
independent of the size of number of background datasets.

\paragraph{Computational complexity.}
Our model needs to perform two main computation tasks for each new
dataset: optimization of the objective function and computation of the
predictive likelihoods.  The optimization step needs to evaluate the
function value, the gradient and maximal element of the gradient; the
computation of the function value has complexity $O(N_q * N_S *
O(\mbox{computing } p({x_i}^q|M^{s_j}))$ while, as discussed in the
previous section, computing the gradient and finding the maximal
element takes linear $O(N_q * N_S)$ time in each iteration if a
fixed step size is used (or if the number of line search evaluations
is restricted below some fixed maximum).  Thus the complete algorithm
takes $ O(I * N_q * N_S * O(\mbox{computing } p({x_i}^q|M^{s_j})) +
O(I*N_q*N_S)$.  Clearly the dominating factor is the computation of
the function value: $O(I *N_q * N_S * O(\mbox{computing }
p({x_i}^q|M^{s_j}))$.

The predictive likelihood for the query sample is computed as its
average probability from $V=1000$ multinomial distributions, estimated
from randomly generated samples coming from the generative process of
the earlier-trained model $M^{s_j}$; this is the standard
\emph{empirical likelihood scheme} discussed in \cite{Li06}.  The
complexity of computing the predictive likelihood for the query
dataset with $G$ features (in our case gene-sets), given a model with
$T$ latent components~\footnote{$T$ is upper bounded by the maximum
  number of samples in a background dataset.} is $O(G*V*T)$. The
total computational complexity is then simply the complexity for
predictive likelihoods times the optimization scheme:
$O(I*N_q*N_S*G*V*T)$.  Since the number of iterations is independent
of the number of background datasets, $N_S$, the complexity is linear
with respect to it, $O(N_S)$, and therefore the model is reasonably
tolerant to the fast growth of public repositories.

In our implementation a single query dataset took about $31$
iterations and $0.15$ seconds on average on an
Intel\textsuperscript{\textregistered} Core (TM) i7 CPU @ 2.93GHz, to
find the optimal weight vector.

\section{Results}

\begin{figure*}[h]
\centering{
\includegraphics{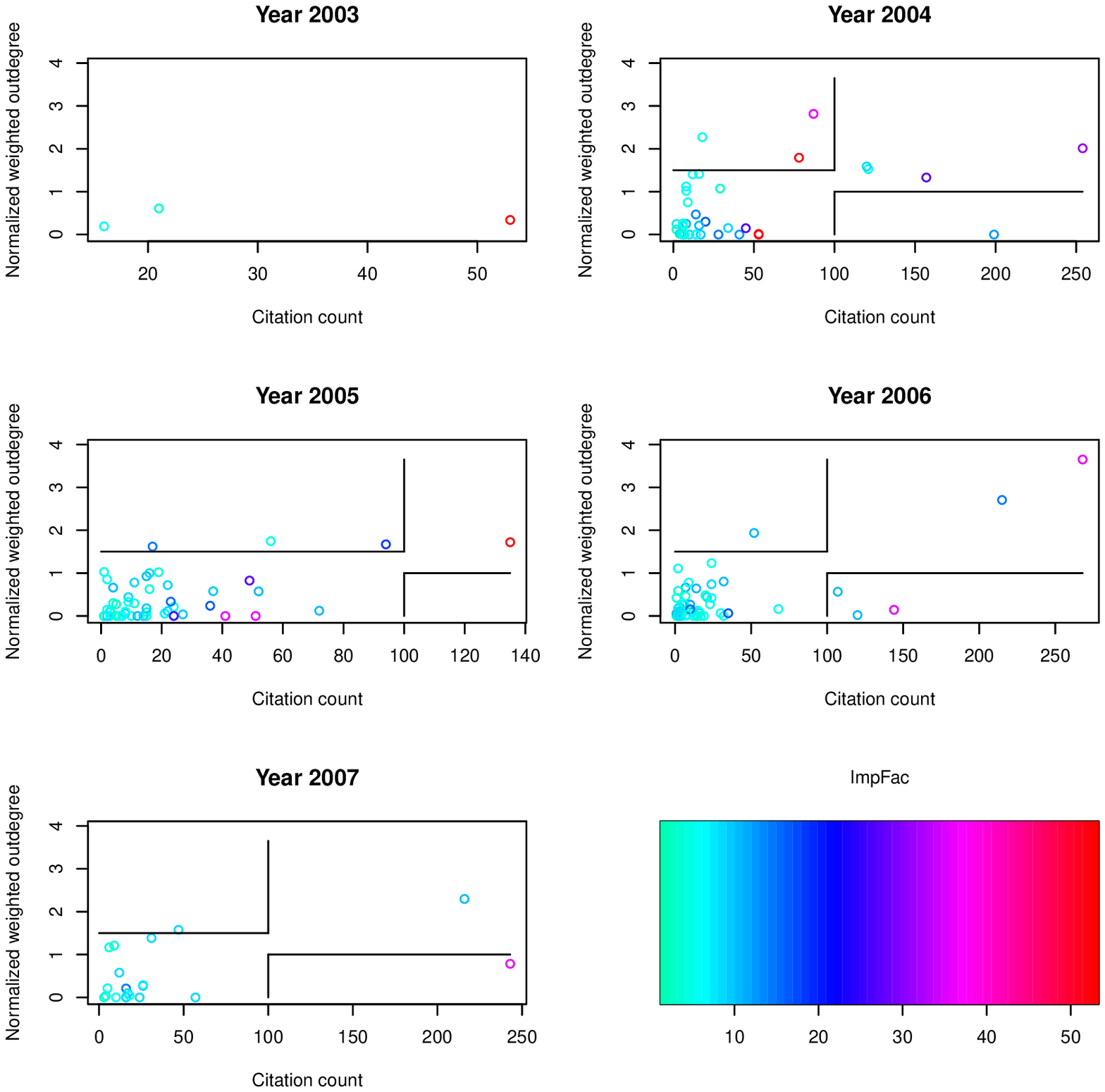}
\caption{Stratified data-driven prediction of usefulness of datasets
  vs. their citation counts. Black solid lines mark the boundary
  for potentially interesting datasets; the boundaries are set to hold the 
  same percentiles of data as in Fig. 3 in the main paper. 
  \emph{ImpFac} stands for Impact Factor of
  the publication venue.\label{fig:scatter_plot_stratified}}}
\end{figure*}

\paragraph{Normalization of citation counts.}  Older datasets tend to
have higher citation counts and outdegrees; in Fig. 3 we removed this
bias by a normalization technique, and identified interesting datasets having very low citation counts and very high outdegrees or vice
versa. To verify the analysis results (identified datasets) are not
artifacts of the normalization, we reanalyzed the original data
without applying citation count normalization.
Fig.~\ref{fig:scatter_plot_stratified} shows the result as stratified
subfigures plotted for each year separately.  The datasets identified
using the original citation counts (top-left and bottom-right corners
in each subfigure of Fig.~\ref{fig:scatter_plot_stratified}) are an
exact match with the datasets identified after normalization.

\begin{figure*}[h]
\centering
\includegraphics{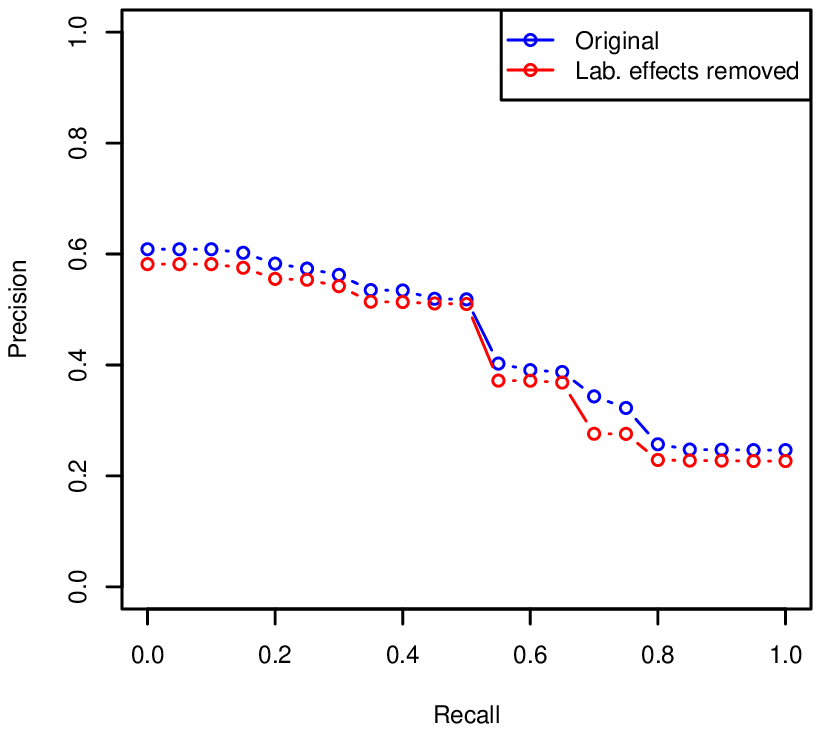}
\caption{Removal of laboratory effects changes the retrieval
  performance only slightly, as measured by the precision-recall curves. 
  \emph{Original:} Replicated from Fig. 1 of the main paper;
  \emph{Lab. effects removed:} all retrieval results from the same laboratory
  as the query data have been discarded.
\label{fig:PR_labEffect}}
\end{figure*}

\paragraph{Retrieval performance after discounting for the laboratory
  effect.} In microarray experiments laboratory effects are known to be
strong \cite{Zilliox07}. The 206 datasets studied were generated from
163 laboratories. The top laboratory was responsible for 7 datasets,
whereas 142 laboraties only contributed a single set. To test how much
the laboratory effects have affected our results, we discarded all
retrieved results from the same laboratory as the query set. The
original precision-recall curve and the corrected curve are in
Figure~\ref{fig:PR_labEffect}; the mean average precision dropped to
$0.44$ from $0.42$. The small change in performance shows that our
result is mainly due to other effects captured by the model than
the laboratory effect.

\paragraph{Quantitative comparison of data-driven results against the
   citation patterns.} Of the 23 direct citation links, eleven were
also found by our model as having a non-zero edge weight. Six links
could not have been found; five of them are citations by papers having
very small datasets ($<10$), which we had considered to be too small
to act as queries while one citation link is between datasets released
on the same date. Of the remaining six links not observed in our
model, two are the cross-cluster citations that are not due to
biological similarity of the datasets as discussed in the paper;
two cell line datasets about multiple myeloma and large cell lymphoma
(GSE6205 and GSE6184 respectively) cite a leukemia study (GSE2113)
where plasma cells are profiled; one dataset measuring HIV
infection from T cells (GSE6740) cites a very small dataset about
thymocyte, the limited size of the set reducing its corresponding
model's relative capability to explain other sets compared to large
datasets in the collection. Finally, E-TABM-26, a prostate cancer study,
cites E-MEXP-156 (a study about tumorigenic and nontumorigenic Human
Embryonic Kidney Cells) in the context of cell apoptosis of cancer
cells in general.

\begin{figure*}[h]
\centering
\includegraphics{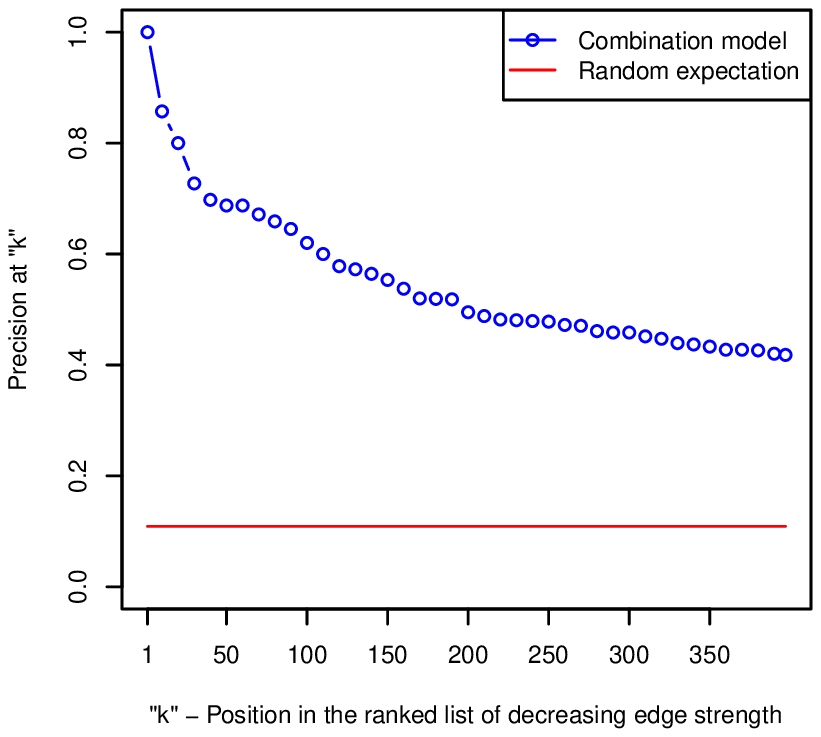}
\caption{Overlap of data-driven recommendations with the actual
  citation graph: Precision $@ k$ for top edges that explain more than
  $2.5\%$ variation.  The gold standard is the extended citation graph
  which is built as the union of edges from 1) the original
  directed graph, 2) between any two articles that are cited together
  by some other article and 3) between any two articles that have at
  least one common reference.
\label{fig:citgraph_comparison}}
\end{figure*}

Vice versa, we evaluated the top-weight data-driven edges against
direct and indirect citation links and found a favorable, non-random
overlap. For this we use a modern standard metric in information
retrieval called precision $@ k$ which measures precision at each
position $k$ in the ranked list of top retrieval results from a search
engine.  In our case the results are the ranked list of inferred edges
in descending order of their edge strengths.  The citation patterns
were used as the gold standard for existence of an edge between two
datasets using their corresponding publication information.
Fig.~\ref{fig:citgraph_comparison} shows that the precision $@ k$ is
reasonably high; for the top two hundred data-driven edges (\emph{i.e.},
$k=200$) the value is 0.5.

\paragraph{Densely connected set of experiments in the relevance network.}
For each of the top 20 strongest edges in the relevance network
(listed in Table~\ref{tab:top20links}) we searched for associated
cliques, \emph{i.e.}, connections within neighbors of the edge, where the
clique size is at least three. We found seven cliques; four of them
(breast cancer, leukocytes, human immune T helper cells and
developmental stages of thymocytes) are described in the main text.
The remaining three are an adenocarcinoma, a brain tissue and a
skeletal muscle clique.  The first clique is among GSE4824, GSE5258
and GSE6914; all three datasets are heterogeneous collections of
different cancerous tissues where majority of the samples are cell
line profiles from either lung or breast adenocarcinoma. The second
clique is among GSE1297, GSE5392 and E-MEXP-114 that profile normal
and diseased brain tissue. The last clique is among all skeletal
muscle experiments of the collection where the strongest edge is
between GSE3307 and GSE6011 that both measure Duchenne muscular
dystrophy sampled from quadriceps muscle tissues, the former also
containing samples form other skeletal muscle diseases.

\begin{figure*}[h]
\centering
\includegraphics{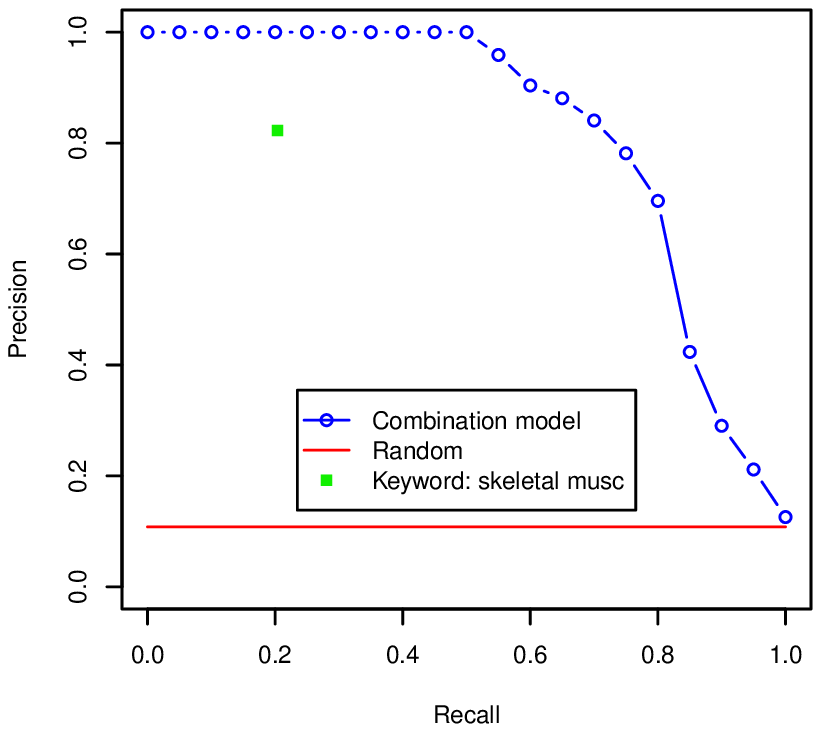}
\caption{Retrieval performance evaluation of the data-driven model
  against keyword search in the skeletal muscle case study. The
  precision-recall curves are averaged across the 16 skeletal muscle
  datasets having at least 10 samples.\label{fig:PR_SM}}
\end{figure*}

\paragraph{Skeletal muscle {retrieval} case study.} 
We lastly present a
case study to illustrate how the model retrieval can support a
researcher in finding relevant data on a specific topic, lessening the
need for laborious manual searches. The topic of interest was gene
expression in skeletal muscle and our database consisted of the human
gene expression atlas~\cite{Lukk12}, which includes eight skeletal
muscle datasets among a total of 206 datasets, plus additional 16
skeletal muscle datasets extracted manually from ArrayExpress
(Table~\ref{tab:smcollection}).  First, we tested how well the
data-driven method retrieved other skeletal muscle datasets when
querying with any single skeletal muscle dataset.  The retrieval
performance across all 16 query datasets was close to optimal, whereas
keyword searches were not as good in the same task
(Fig.~\ref{fig:PR_SM}).  The reason for that was the lack of a
consistent annotation vocabulary in the dataset descriptions; in
particular, different levels of specificity had been used.

Next we looked in more detail into the retrieval results of the
individual queries. For all of them, the retrieval result was
sparse, {\em i.e.}, less than 10\% of the datasets were found to be
relevant to the query (by a non-zero weight).  
We further investigated the ranking of results provided by
the data-driven retrieval model. 
For 12 queries all retrieved results were other skeletal muscle datasets;
results for the remaining queries contained
at least one false positive, and they are summarized in Table~\ref{tab:sm}. 
The top retrieved non-skeletal muscle dataset is the brain tissue 
dataset GSE5392, followed by the kidney dataset GSE781.
Kidney does not have a direct biological connection to skeletal
muscle. It is known that some areas of kidney are rich in blood
vessels, which are lined by smooth muscle. Skeletal muscle, smooth 
muscle and cardiac muscle are the three main muscle types in the human
body. Interestingly, one skeletal muscle datasets, E-MEXP-216, was never retrieved 
by any skeletal muscle query, which suggests
that it is an outlier in some respect. The dataset contains
a combination of human and macacque liver and skeletal muscle
samples. It contains only four human skeletal muscle samples. 

Finally, the internal ranking of skeletal muscle datasets in response
to a particular query dataset seems to be guided to a large extent by
health conditions:

\begin{itemize}
\item {\bf E-GEOD-9397} contains samples annotated with disease status FSHD
  (Facioscapulohumeral muscular dystrophy). The top retrieved dataset
  for that query is E-GEOD-10760, the only other dataset in the
  collection annotated with FSHD.

\item The top retrieved result for {\bf E-GEOD-12648} is {\bf
  E-GEOD-11686} and vice versa. Both of them are neuromuscular
  disorders, hereditary inclusion body myopathy (HIBM) and cerebral
  palsy, and they are the only datasets with these specific diseases
  in our data collection.

\item {\bf E-GEOD-1786} partially contains samples from COPD (Chronic
  Obstructive Pulmonary Disease) subjects; the disease has a muscle
  wasting effect. The top retrieved result is
  E-GEOD-10760; it contains samples from the same muscle type, but
  another disease (FSHD) which also leads to muscle wasting.

\item {\bf E-GEOD-1295} contains samples of the trained and untrained
  muscle of non-young overweight people with prediabetic metabolic
  syndrome.  Half of the samples in E-GEOD-1786 are also from trained
  muscles (of COPD patients and controls, old overweight
  men). E-GEOD-1786 appears at rank 6, and it is the only background
  dataset known to contain trained muscle samples.  Among the top 5
  datasets, 3 are annotated to contain disease samples related to
  weakening of muscles, another one is known to contain old overweight
  samples ({\bf E-GEOD-8441}).
\end{itemize}
In summary, disease conditions and health states of tissue seem to
determine the ranking within datasets of the same tissue (sub)type.

\newpage

\begin{figure*}[h]
\begin{center}
\includegraphics{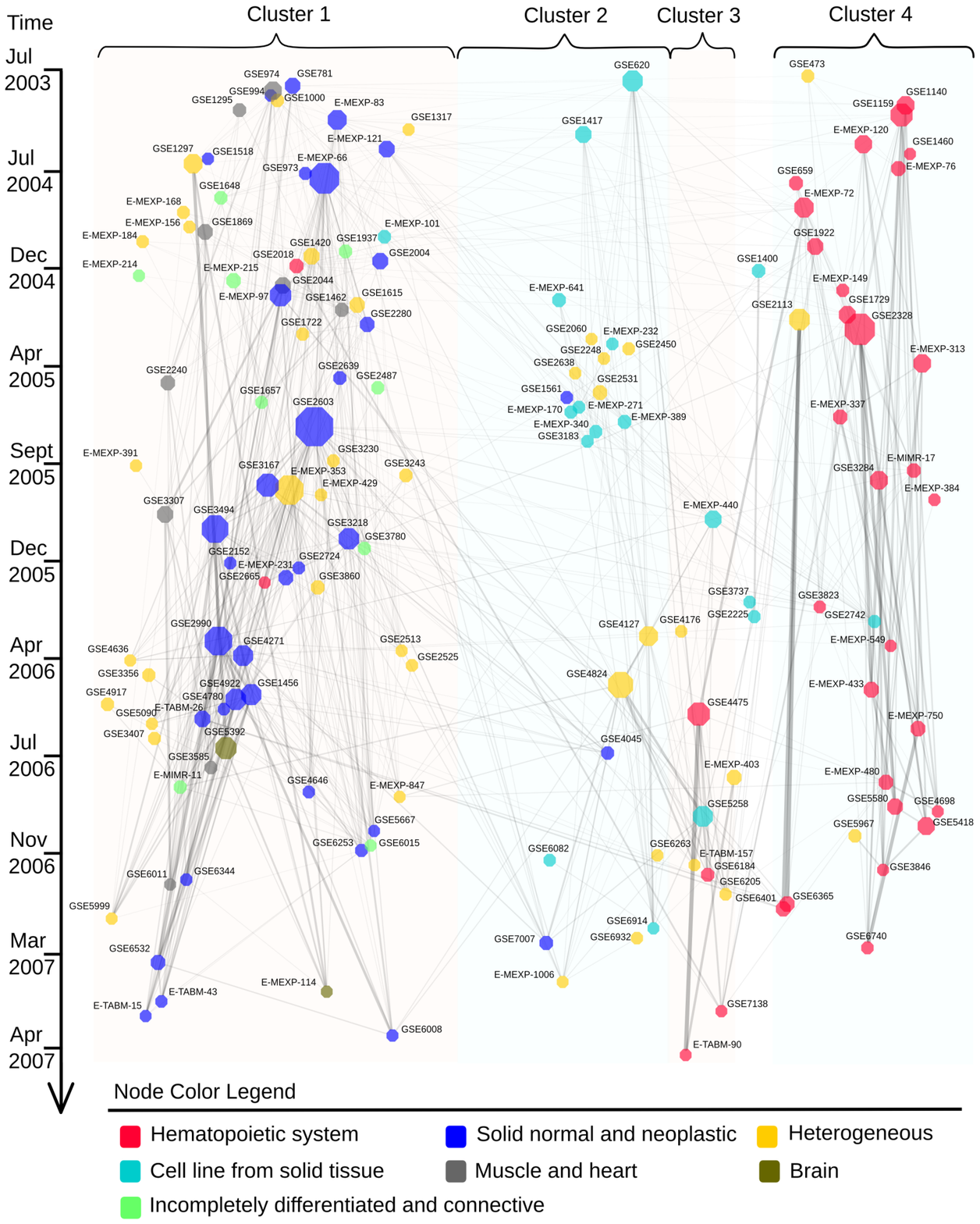}
\end{center}
\caption{Enlarged visualization of the relevance network of datasets
  in the human gene expression atlas (Fig. 2 in the main paper); each
  dataset was used as a query to retrieve earlier datasets. A link
  from an earlier dataset to a later one means the earlier dataset is
  relevant as a partial model of activity in the later dataset. The
  link width is proportional to the normalized relevance weight
  (combination weight $\theta_{j}^q$; only links with
  $\theta_{j}^q\geq 0.025$ are shown, and datasets without links have
  been discarded). The node size is proportional to the estimated
  influence, \emph{i.e.}, the total outgoing weight. Colors: tissue
  types (the six meta tissue types \cite{Lukk12}). The node layout was
  computed from the data-driven network using the cluster-emphasizing
  Sammon mapping scheme. Cytoscape \cite{Shannon03} was used to
  generate the figure.}
\end{figure*}

\begin{table*}
\begin{tabular}{lll}
\hline
Edge & Clique membership \\
\hline
GSE4475 - E-TABM-90 & Does not form a clique\\
GSE2113 - GSE6365 & Does not form a clique\\
GSE2113 - GSE6401 & Does not form a clique\\
E-MEXP-750 - GSE6740 & Part of the CD4+ T helper cell clique\\
GSE1297 - GSE5392 & Part of the brain tissue clique\\
GSE2328 - E-MEXP-433 & Extends the leukocyte clique\\
GSE2990 - GSE6532 & Part of the breast cancer clique\\
GSE2328 - GSE5580 & Part of the leukocyte clique\\
GSE2990 - GSE1456 & Part of the breast cancer clique\\
GSE3494 - GSE2990 & Part of the breast tissue clique\\
GSE2990 - GSE6008 & Does not form a clique\\
GSE4824 - GSE5258 & Part of the adenocarcinoma clique\\
E-MEXP-72 - GSE4475 & Does not form a clique\\
E-MEXP-403 - E-TABM-90 & Does not form a clique\\
GSE3494 - GSE1456 & Part of the breast tissue clique\\
E-MEXP-337 - E-MEXP-549 & Part of the developmental stage of lymphocytes clique\\
GSE3284 - E-MEXP-433 & Part of the leukocyte clique\\
GSE3307 - GSE6011 & Part of the skeletal muscle datasets clique\\
GSE2990 - GSE4922 & Part of the breast cancer clique\\
GSE2044 - GSE3307 & Part of the skeletal muscle datasets clique\\
\hline
\end{tabular}
\caption{Top $20$ strongest edges in the relevance
  network. \label{tab:top20links}} 
\end{table*}

\begin{table*}
\begin{tabular}{llll}
E-CBIL-30 & E-GEOD-10760 & E-GEOD-11686 & E-GEOD-11971 \\
E-GEOD-12648 & E-GEOD-1295 & E-GEOD-1786 & E-GEOD-3307 \\
E-GEOD-4667 & E-GEOD-474 & E-GEOD-6011 & E-GEOD-7146 \\
E-GEOD-8441 & E-GEOD-9105 & E-GEOD-9397 & E-GEOD-9676 \\
\end{tabular}
\caption{ArrayExpress accession numbers of 16 skeletal muscle datasets
  used in the retrieval case study in addition to the human gene
  expression atlas~\cite{Lukk12}. All datasets were measured with the
  human genome platform HG-U133A, the same used in the
  atlas.\label{tab:smcollection}}
\end{table*}

\begin{table*}
\caption{Skeletal muscle queries with at least one retrieved non-skeletal muscle dataset, 
sorted according to decreasing precision.\label{tab:sm}}
\begin{tabular}{lrrrl}
\hline
Query & No. of retrieved datasets & No. of retrieved skeletal muscle datasets  & Top false positive dataset (rank) \\
\hline
E-GEOD-4667 & 20 & 18  & GSE5392 (15) \\
E-GEOD-11686 & 26 & 22  & GSE5392 (22) \\
E-GEOD-7146 & 25 & 21  & GSE781 (22) \\
E-GEOD-11971 & 25 & 19  & GSE5392 (15) \\
\hline
\end{tabular}
\end{table*}

\end{article}